\documentclass[reprint,aps,prl,twocolumn,superscriptaddress]{revtex4-1}
\usepackage{graphicx}
\usepackage{dcolumn}
\usepackage{bm}
\usepackage{amsmath}
\usepackage[english]{babel}
\usepackage[utf8]{inputenc}
\begin{document}

\title{Quantum transport properties of industrial $^{28}$Si/$^{28}$SiO$_2$}

\author{D. Sabbagh}
\affiliation{QuTech and Kavli Institute of Nanoscience, Delft University of Technology, PO Box 5046, 2600 GA Delft, The Netherlands}
\author{N. Thomas}
\affiliation{Intel Corporation, Technology and Manufacturing Group, Hillsboro, OR 97124, USA}
\author{J. Torres}
\affiliation{Intel Corporation, Technology and Manufacturing Group, Hillsboro, OR 97124, USA}
\author{R. Pillarisetty}
\affiliation{Intel Corporation, Technology and Manufacturing Group, Hillsboro, OR 97124, USA}
\author{P. Amin}
\affiliation{Intel Corporation, Technology and Manufacturing Group, Hillsboro, OR 97124, USA}
\author{H. C. George}
\affiliation{Intel Corporation, Technology and Manufacturing Group, Hillsboro, OR 97124, USA}
\author{K. Singh}
\affiliation{Intel Corporation, Technology and Manufacturing Group, Hillsboro, OR 97124, USA}
\author{A. Budrevich}
\affiliation{Intel Corporation, Technology and Manufacturing Group, Hillsboro, OR 97124, USA}
\author{M. Robinson}
\affiliation{Intel Corporation, Technology and Manufacturing Group, Hillsboro, OR 97124, USA}
\author{D. Merrill}
\affiliation{Intel Corporation, Technology and Manufacturing Group, Hillsboro, OR 97124, USA}
\author{L. Ross}
\affiliation{Intel Corporation, Technology and Manufacturing Group, Hillsboro, OR 97124, USA}
\author{J. Roberts}
\affiliation{Intel Corporation, Technology and Manufacturing Group, Hillsboro, OR 97124, USA}
\author{L. Lampert}
\affiliation{Intel Corporation, Technology and Manufacturing Group, Hillsboro, OR 97124, USA}
\author{L. Massa}
\affiliation{QuTech and Kavli Institute of Nanoscience, Delft University of Technology, PO Box 5046, 2600 GA Delft, The Netherlands}
\author{S. V. Amitonov}
\affiliation{QuTech and Kavli Institute of Nanoscience, Delft University of Technology, PO Box 5046, 2600 GA Delft, The Netherlands}
\author{J. M. Boter}
\affiliation{QuTech and Kavli Institute of Nanoscience, Delft University of Technology, PO Box 5046, 2600 GA Delft, The Netherlands}
\author{G. Droulers}
\affiliation{QuTech and Kavli Institute of Nanoscience, Delft University of Technology, PO Box 5046, 2600 GA Delft, The Netherlands}
\author{H. G. J. Eenink}
\affiliation{QuTech and Kavli Institute of Nanoscience, Delft University of Technology, PO Box 5046, 2600 GA Delft, The Netherlands}
\author{M. van Hezel} 
\affiliation{URENCO Stable Isotopes, PO Box 158, 7600 AD, Almelo, The Netherlands}
\author{D. Donelson} 
\affiliation{Air Liquide Advanced Materials North Branch, NJ 08876, USA}
\author{M. Veldhorst}
\affiliation{QuTech and Kavli Institute of Nanoscience, Delft University of Technology, PO Box 5046, 2600 GA Delft, The Netherlands}
\author{L. M. K. Vandersypen}
\affiliation{QuTech and Kavli Institute of Nanoscience, Delft University of Technology, PO Box 5046, 2600 GA Delft, The Netherlands}
\author{J. S. Clarke}
\affiliation{Intel Corporation, Technology and Manufacturing Group, Hillsboro, OR 97124, USA}
\author{G. Scappucci}
\email{g.scappucci@tudelft.nl}
\affiliation{QuTech and Kavli Institute of Nanoscience, Delft University of Technology, PO Box 5046, 2600 GA Delft, The Netherlands}

\date{\today}

\begin{abstract}
We investigate the structural and quantum transport properties of isotopically enriched $^{28}$Si/$^{28}$SiO$_2$ stacks deposited on 300 mm Si wafers in an industrial CMOS fab. Highly uniform films are obtained with an isotopic purity greater than 99.92\%. Hall-bar transistors with an equivalent oxide thickness of 17 nm are fabricated in an academic cleanroom. A critical density for conduction of $1.75\times10^{11}$ cm$^{-2}$ and a peak mobility of 9800 cm$^2$/Vs are measured at a temperature of 1.7 K. The $^{28}$Si/$^{28}$SiO$_2$ interface is characterized by a roughness of $\Delta=0.4$ nm and a correlation length of $\Lambda=3.4$ nm. An upper bound for valley splitting energy of 480 $\mu$eV is estimated at an effective electric field of 9.5 MV/m. These results support the use of wafer-scale $^{28}$Si/$^{28}$SiO$_2$ as a promising material platform to manufacture industrial spin qubits. 
\end{abstract}

\maketitle

Enrichment of the spin-zero $^{28}$Si isotope drastically reduces spin-bath decoherence in silicon \cite{PhysRevLett.105.187602, witzel2006quantum} and has enabled solid state spin qubits with extremely long coherence \cite{muhonen2014storing, veldhorst_two-qubit_2015} and high control fidelity \cite{ veldhorst_addressable_2014, yoneda_quantum-dot_2018, huang_fidelity_2018}. The limited availability of isotopically enriched $^{28}$Si in industrially adopted forms \cite{itoh_isotope_2014}, however, was previously thought to be a major bottleneck to leverage CMOS technology for manufacturing qubits with the quality and in the large numbers required for fault tolerant quantum computation \cite{vandersypen2017interfacing,li2018crossbar}. Recently, isotopically enriched silane ($^{28}$SiH$_{4}$) has been employed in a pre-industrial CMOS facility to deposit high quality $^{28}$Si epi-wafers \cite{mazzocchi201899}. Crucially, an industrial supply of $^{28}$SiH$_{4}$ has been established and silicon quantum dots were fabricated on a wafer-scale $^{28}$Si/$^{28}$SiO$_2$ stack grown in an industrial manufacturing CMOS fab \cite{petit2018spin}. In these quantum dots, a single-electron spin lifetime of 2.8 ms was obtained at a temperature of 1.1 K and weak charge noise was measured, pointing to a promising material platform for qubit operation at elevated temperatures.

Studies devoted to $^{28}$Si quantum dots, however, tend to discuss only marginally the structural properties of the originating $^{28}$Si/$^{28}$SiO$_2$ material stack and the electrical transport in the associated two-dimensional electron gas (2DEG). In this paper we provide structural characterization of the same industrial $^{28}$Si wafer used for quantum dots in Ref. \cite{petit2018spin} and assess the disorder properties of the critical $^{28}$Si/$^{28}$SiO$_2$ interface. By investigating the quantum transport properties of Hall-bar transistors, we extract key material metrics such as carrier mobility, critical density, interface roughness, interface correlation length, and valley splitting energy. Electron mobility is typically used as a figure of merit to assess the quality of the semiconductor/oxide interface. However, peak mobility is measured at high electron density, where screening effects are relevant \cite{tracy_2009,kim_lyon_2017}. The critical density, instead, indicates the minimum density required to establish metallic conduction by overcoming electron trapping at the oxide interface. As such, the critical density is a complementary metric to the mobility and characterizes the interface disorder at low densities, where quantum devices typically operate. Overall, large mobility and small critical density indicate material uniformity and low disorder at the confining interfaces. These properties are beneficial for obtaining reproducible quantum dots at intended locations on the substrate. Valley splitting quantifies the energetic separation between the ground state used for computation and the lowest excited state. A sharp flat interface is required to achieve large splitting energy, which is beneficial for qubit operation \cite{culcer2010interface,yang2013spin}. The results reported in this work indicate a low disorder environment at the $^{28}$Si/$^{28}$SiO$_2$ interface and potential to achieve large valley splitting, supporting the industrial integration of spin qubits on wafer-scale $^{28}$Si.

\begin{figure*}%
	\includegraphics[width=170mm]{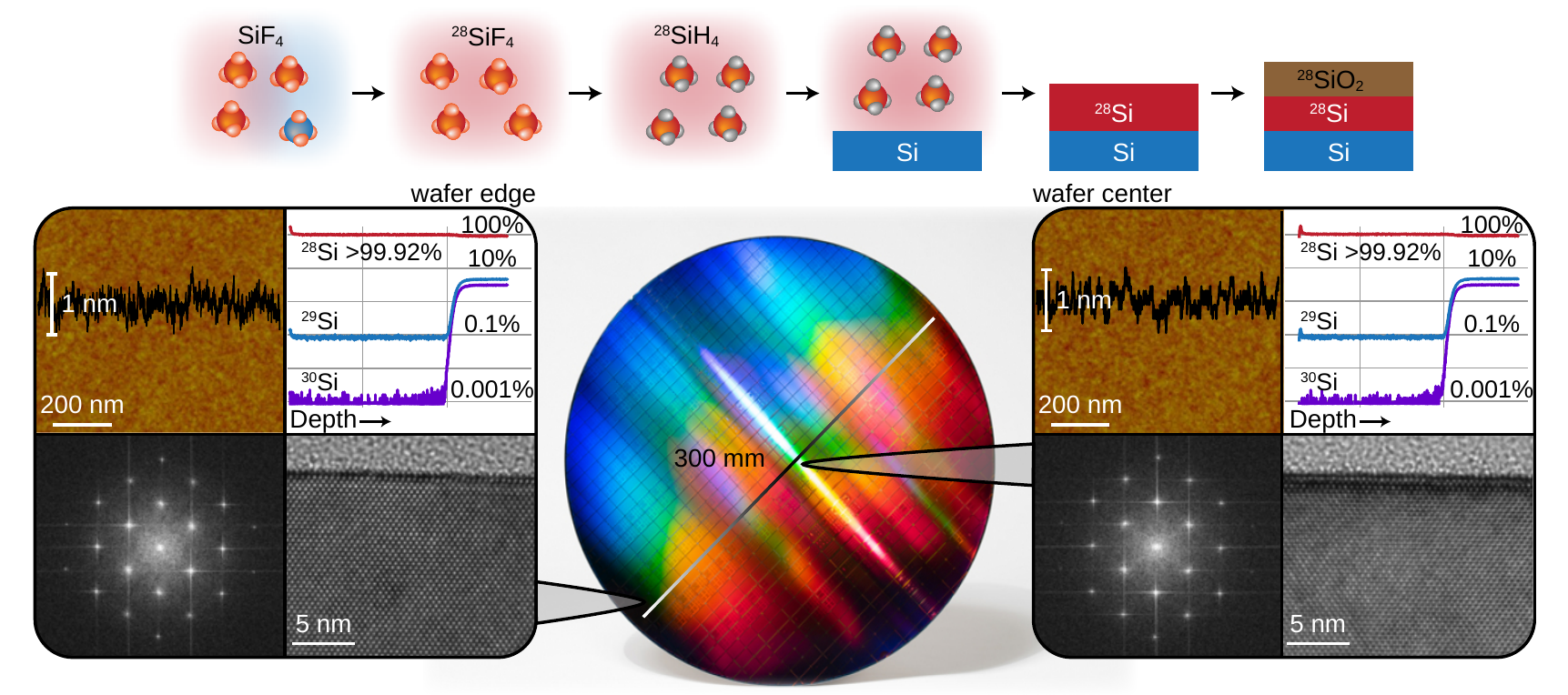}%
	\caption{Central panel: a 300 mm Si(100) wafer with epitaxial $^{28}$Si. Top panel: supply chain for $^{28}$SiH$_4$ gas precursor, starting from a natural SiF$_4$ with small concentration of $^{29}$Si (blue) which is enriched in $^{28}$Si (red). The gas is reduced to obtain $^{28}$SiH$_4$ and then used to deposit the $^{28}$Si epilayer followed by thermal oxidation. Atoms of F and H are depicted in orange and gray, respectively. Side panels show the material characterization at the center and at the edge of the as-grown wafer (right and left panels, respectively). The characterization includes (from top left, clockwise): atomic force microscopy of the smooth surface including a line scan showing the vertical profile with maximum excursion of 1 nm; compositional analysis (depth range of 160 nm) by secondary ion mass spectroscopy of isotopes $^{28}$Si (red), $^{29}$Si (blue), $^{30}$Si (purple); high-resolution transmission electron microscope image of the $^{28}$Si/$^{28}$SiO$_2$ interface; electron diffraction patterns with sharp and equally spaced peaks.}
\label{fig:MAT}
\end{figure*}

The schematics in Fig.\ \ref{fig:MAT} illustrate the key steps in the supply chain of isotopically enriched precursors for wafer-scale epitaxial growth of $^{28}$Si. A silicon-tetrafluoride gas (SiF$_4$) with natural abundance of $^{28}$Si of 92.23\% is isotopically enriched in $^{28}$Si to a concentration greater than 99.92\% by centrifuge separation. The $^{28}$SiF$_4$, with a residual $^{29}$Si concentration of 0.08\%, is then reduced to obtain high purity $^{28}$SiH$_4$. $^{28}$SiH$_4$ gas cylinders have been installed for use in a state-of-the-art chemical vapour deposition tool of a 300 mm fabrication line to deposit $^{28}$Si epilayers. Maintaining the chemical purity of gas precursors throughout the supply chain is crucial to obtain a low-disorder $^{28}$Si/$^{28}$SiO$_2$ stack. The growth process starts with the deposition of 1 $\mu$m of intrinsic natural Si on a high-resistivity 300 mm Si(100) wafer followed by a 100-nm-thick intrinsic $^{28}$Si epilayer. The wafer is then thermally processed at high temperature for the formation of a high quality 10-nm-thick $^{28}$SiO$_2$ layer.

In Fig.\ \ref{fig:MAT} we compare morphology and composition of the grown stack at the center and the edge of the 300 mm wafer. No difference in surface or interface roughness, composition, and purity could be observed across the wafer, indicating a uniform film deposition.  Atomic force microscopy shows a uniform and near defect-free surface with a root mean square surface roughness of 0.2 nm. Secondary ion mass spectroscopy of isotopes $^{28}$Si, $^{29}$Si, and $^{30}$Si shows a high purity film with a residual concentration of non-zero-spin nuclei $^{29}$Si reduced from $4.76\%$ in the Si buffer to $0.08\%$ in the purified epilayer, demonstrating that the precursor purity has been preserved during the deposition process. The concentration of common background contaminants C and O is below the detection limit of $4\times10^{17}$ cm$^{-3}$ and $1\times10^{18}$ cm$^{-3}$, respectively. High-resolution transmission electron microscopy shows that no dislocations or stacking faults are visible in the epilayer. Moreover, the $^{28}$Si/$^{28}$SiO$_2$ interface is flat down to 1-2 atomic layers over distances (${\simeq}200$nm) that are larger than the typical size of Si quantum dot spin qubits. The sharpness of the interface, the negligible density of defects in the lattice, and the associated electron diffraction pattern highlight the film quality and the good control over the growth process attained in a manufacturing CMOS fab.

\begin{figure}
  \includegraphics[width=85mm]{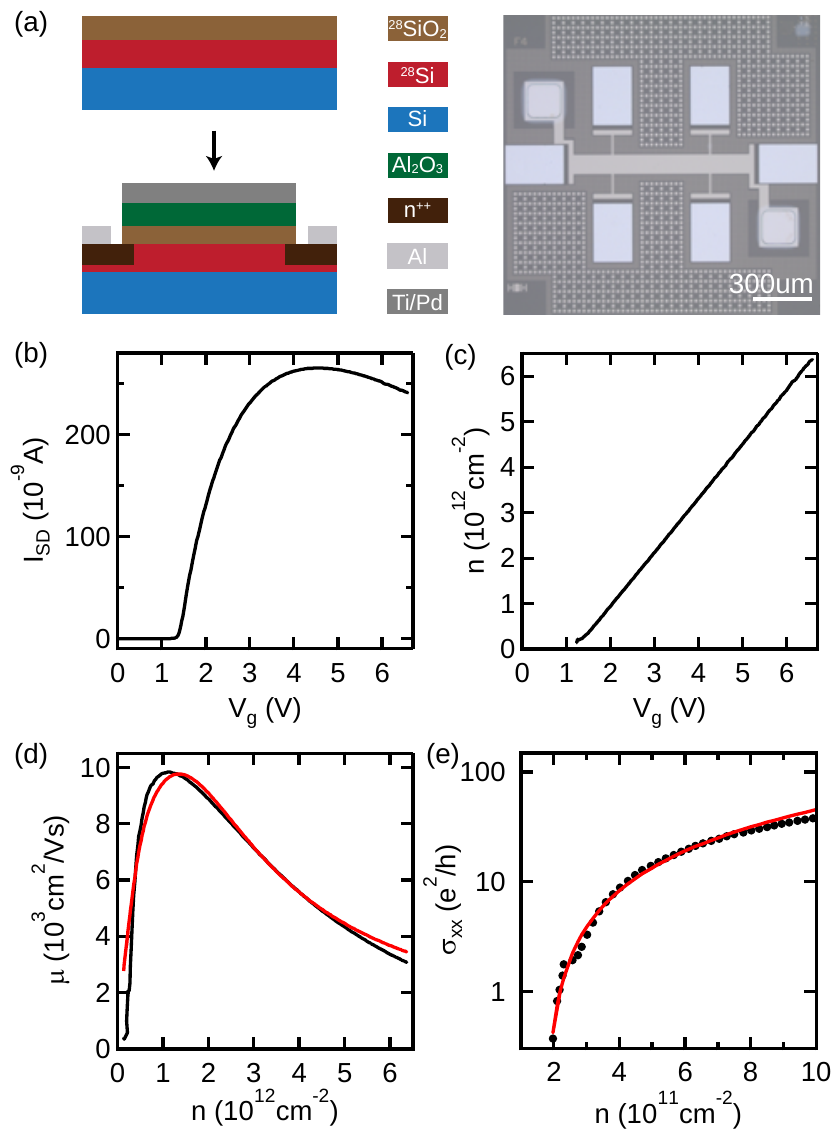}
  \caption{(a) Schematics of the Hall-bar device fabrication, starting from the 300 mm $^{28}$SiO$_2$/$^{28}$Si/Si(100) stack followed by coupon-size processing. The optical micrograph of the final device shows the multi-terminal geometry used for Hall measurements. (b) Source-drain current $I_{SD}$ measured as a function of top gate voltage $V_g$ at $T$ = 1.7 K. (c) Linear relationship between 2DEG Hall density $n$ and top gate voltage $V_g$. (d) Channel mobility $\mu$ measured as a function of $n$ (black) and corresponding calculation (red) including scattering from charged impurities and from interface roughness. (e) 2DEG conductivity in the low density range (black) and fit to percolation theory (red).}
  \label{fig:CLA}
\end{figure}

Moving on to device fabrication, Fig.\ \ref{fig:CLA}(a) shows schematics and optical micrograph of a MOS transistor shaped in a Hall-bar geometry to investigate the magnetotransport properties of the 2DEG at the $^{28}$Si/$^{28}$SiO$_2$ interface. The device was fabricated in an academic cleanroom environment, starting from coupon-sized samples diced from the original $^{28}$Si/$^{28}$SiO$_2$ 300 mm wafer. We employ e-beam lithography and lift-off additive techniques to resemble the process flow used to fabricate quantum dots as in Ref. \cite{petit2018spin}. Highly doped n$^{++}$ regions are obtained by P-ion implantation followed by 30 s of activation anneal at 1000$^\circ$C in N$_2$ environment. Multiple ohmic contacts are deposited on the implant regions by e-beam evaporation of Al. An additional Al$_2$O$_3$ layer was deposited by atomic layer deposition at 300$^\circ$C, so that the $^{28}$Si/$^{28}$SiO$_2$ interface undergoes similar processing as in the fabrication of multi-layer gate-defined qubits \cite{petit2018spin}. The resulting dielectric stack has an equivalent oxide thickness (EOT) of 17 nm. A Ti/Pd top gate is deposited to define a Hall-bar geometry with a 100-$\mu$m-wide and 500-$\mu$m-long central region. The last processing step is a forming gas anneal at 400$^\circ$C to reduce the damage induced by e-beam lithography \cite{nordberg2009enhancement,kim_lyon_2017}. 

The electrical characterization of the device was performed at a temperature of 1.7 K using standard four-terminal low frequency lock-in techniques with a constant source-drain excitation voltage of 1 mV. Longitudinal ($\rho_{xx}$) and transverse ($R_{xy}$) resistivity were measured as a function of carrier density - controlled by the top gate - and external perpendicular magnetic field $B$. The Hall carrier density $n$ and the electron mobility $\mu$ are calculated using the relationships $R_{xy}=(ne)^{-1}B$ and $\mu=(ne\rho_0)^{-1}$, respectively, where $e$ is the elementary charge and $\rho_0=\rho_{xx}(B=0$ T).

A DC voltage applied to the top gate ($V_g$) accumulates a 2DEG at the $^{28}$Si/$^{28}$SiO$_2$ interface which conducts above a turn-on voltage of $V_{to}$ = 1.22 V, as shown in Fig.\ \ref{fig:CLA}(b). For values below $V_{to}$ no current flow is observed in the device up to a temperature of $T= 23$ K, confirming the insulating behavior of the intrinsic $^{28}$Si film at low temperature. For $V_g > V_{to}$ (Fig.\ \ref{fig:CLA}(c)) we measure a linear increase in the Hall density $n$ as a function of $V_g$. The experimental capacitance $C$ = $e$ $dn/dV_g=0.19$ $\mu$F/cm$^2$ matches within 5\% the expected value for the given EOT. Upon multiple sweeps of $V_g$ no hysteresis is observed and the same values of $V_{to}$ and $C$ are measured, indicating a stable potential landscape at the oxide interface.

The experimental and theoretically calculated density-dependent mobility curves are shown in Fig. \ref{fig:CLA}(d). Above a critical density, required to establish metallic conduction in the 2DEG, the mobility rises sharply due to screening from charged impurity Coulomb scattering \cite{ando1982electronic,gold_1985,gold1986scattering,kruithof1991temperature}. A peak mobility of 9800 cm$^2$/Vs is reached at $n = 1.13\times10^{12}$ cm$^{-2}$, corresponding to a mean free path of 120 nm. Beyond, the mobility drops due to surface roughness scattering at the $^{28}$Si/$^{28}$SiO$_2$ interface \cite{ando1982electronic,gold1986scattering,kruithof1991temperature}. The calculated scattering-limited mobility takes into account a scattering charge density at the semiconductor/oxide interface and an exponential autocorrelation function form of the interface roughness \cite{gold1986scattering,kruithof1991temperature,tracy_2009}. A good match is obtained for a scattering charge density of $4.65\times10^{10}$ cm$^{-2}$, an interface roughness of $\Delta=0.4$ nm, and an interface correlation length of $\Lambda=3.4$ nm. $\Delta$ describes the interface root-mean-square height fluctuations, $\Lambda$ is the lateral distance over which the fluctuations are correlated. The interface roughness is compatible with the morphology investigation by transmission electron microscopy reported in Fig.\ \ref{fig:MAT}.
\begin{table*}[!htbp]
\caption{Comparison of material stack, device characteristics, and low-temperature electrical transport properties in Si Hall-bar transistors obtained by different groups. We also indicate whether or not quantum dots have been fabricated on the same wafer.}
\label{table:comparison}
\begin{ruledtabular}
\centering
\begin{tabular}{lcccccccc}

Reference & This work & Ref. \cite{rochette2017dot28Si} & Ref. \cite{shankar_lyon2010spin}& Ref. \cite{nordberg2009enhancement} & Ref. \cite{kim_lyon_2017} & Ref. \cite{kim_lyon_2017} & Ref. \cite{tracy_2009} & Ref. \cite{tracy_2009}\\
\hline
Channel material  & 99.92\% $^{28}$Si-epi & 99.95\% $^{28}$Si-epi & $^{28}$Si-epi & Si bulk & Si bulk & Si bulk & Si bulk & Si bulk\\
Equivalent oxide thickness (nm) & 17 & 35 & 110 & 62 & 35 & 35 & 30 & 10\\
E-beam exposure/forming gas & Yes & N/A & N/A & Yes & Yes & No & No & No \\
Mobility (10$^3$ cm$^2$/Vs) & 9.8 & 11.6 & 14 & 8.3 & 14 & 23 & 15 & 10\\
Critical density (10$^{11}$ cm$^{-2}$) & 1.75 & 1.60 & N/A & N/A & 0.95\footnote{Percolation transition density extrapolated at $T = 0$.} & 0.83\footnotemark[1] & 1.04\footnotemark[1] & N/A \\
Quantum dot fabrication & Yes \cite{petit2018spin} & Yes & No & Yes & No & No & No & No\\

\end{tabular}
\end{ruledtabular}
\end{table*}

The critical density is extracted from a percolation fit of the density-dependent conductivity (Fig.\ \ref{fig:CLA}(e)) $\sigma_{xx}\sim (n-n_p)^p$ \cite{kim_lyon_2017,tracy_2009}, where $n_p$, $p$ are the percolation
transition density and exponent, respectively. By fixing $p=1.31$, as expected in a 2D system, we estimate $n_p=1.75\times 10^{11}$ cm$^{-2}$ at $T=1.7$ K. Previous studies have shown that $n_p$ decreases with decreasing temperature \cite{kim_lyon_2017,tracy_2009}, therefore the obtained value of $1.75\times 10^{11}$ cm$^{-2}$ sets an upper bound for the critical density in the temperature regime at which qubits are typically operated ($T\leq 100$ mK). 

Both the mobility and critical density obtained in wafer-scale isotopically enriched $^{28}$Si/$^{28}$SiO$_2$ stack are qualitatively comparable to the values previously reported for high-mobility Si MOSFETs at low temperatures \cite{rochette2017dot28Si,shankar_lyon2010spin,kim_lyon_2017,nordberg2009enhancement,tracy_2009} (see Table \ref{table:comparison}). In drawing a meaningful comparison with the data reported in the literature, the reader should consider samples with similar EOT and device process flow. In fact, the mobility is known to be higher in devices with thicker oxide \cite{ando1982electronic,yagi1978oxide} and degrades upon device exposure to electron-beam \cite{nordberg2009enhancement,kim_lyon_2017}. 

\begin{figure}[!h]
  \includegraphics[width=\linewidth]{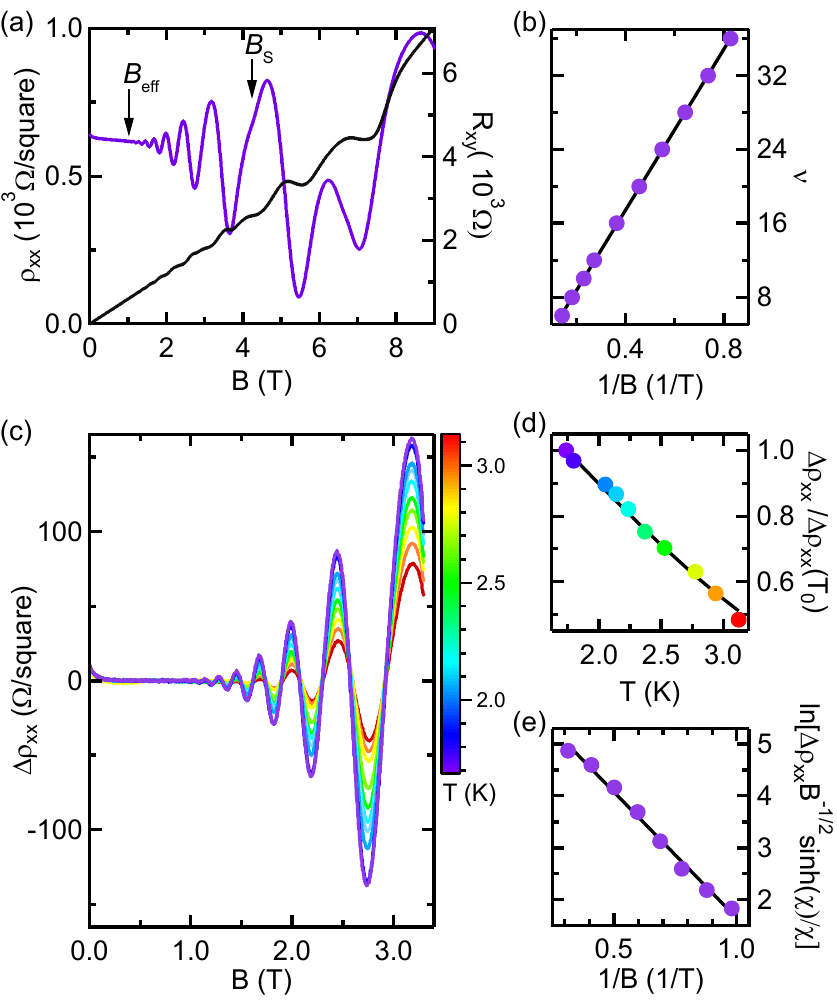}
  \caption{(a) Longitudinal ($\rho_{xx}$, purple) and transverse ($R_{xy}$, black) resistivity at $n=1.05\times10^{12}$ cm$^{-2}$ and $T=1.7$ K. The arrows indicate the magnetic field at which SdH oscillations and Zeeman spin-splitting are resolved. (b) Linear relationship between the filling factors $\nu$ and the inverse of magnetic field $B$. The solid line is the linear fit from which $n_{SdH}$ is calculated. (c) Temperature dependence of the SdH oscillations amplitude in the range $T=1.7{-}3.1$ K, after polynomial background subtraction. (d) $\Delta\rho_{xx}$ at $B=3.18$ T as a function of $T$, normalised at the value at $T=1.7$ K. The solid line is the fit used to extract $m^*$. (e) Dingle plot at $T=1.7$ K, considering the 8 most resolved oscillation maxima. The solid line is the linear fit used to extract $\tau_q$.}
  \label{fig:QUANT}
\end{figure}

Transport characterization at high magnetic field (Fig.\ \ref{fig:QUANT}) allows the measurement of effective mass $m^*$ and quantum lifetime $\tau_q$, from which we estimate an upper bound for the valley splitting energy and the $g$-factor. In Fig.\ \ref{fig:QUANT}(a) we report the longitudinal magnetoresistivity at a density $n=1.05\times10^{12}$ cm$^{-2}$ which corresponds to an effective electric field of 9.5 MV/m.
Shubnikov-de Haas (SdH) oscillations are observed, with minima aligned to quantum Hall plateaus in $R_{xy}$. SdH oscillations start at $B_{eff}= 1$ T and spin degeneracy is resolved at $B_S= 4.3$ T, corresponding to the even filling factor $\nu=10$. Figure \ref{fig:QUANT}(b) shows the filling factor progression against $1/B$. High mobility and density allow to resolve filling factors up to $\nu=36$, with fourfold periodicity at low magnetic field due to spin and valley degeneracy and twofold periodicity beyond $B_S$. We do not observe odd filling factors, indicating that twofold valley degeneracy is not resolved under these measurement conditions. From the linear filling factor progression we extract a density $n_{SdH}=1.06\times10^{12}$ cm$^{-2}$. The agreement between the Hall density $n$ and $n_{SdH}$ indicates that only one high-mobility subband contributes to electrical transport, confirming the high quality $^{28}$Si epitaxy.

The transverse effective mass $m^*$ of the high mobility carriers is calculated from the damping of the SdH oscillations with increasing temperature, described by the relation
\cite{bauer_1972,isihara_1986,coleridge_1991,celik_2011}
\begin{equation}\label{eq:effmass}
\dfrac{\Delta\rho_{xx}(T,B)}{\Delta\rho_{xx}(T_0,B)}=\dfrac{T\sinh\chi_0}{T_0\sinh\chi} \: ,
\end{equation}
where $\Delta\rho_{xx}$ is the SdH oscillation amplitude after polynomial background subtraction, $\chi=2\pi^2k_BT/\hbar\omega_c$, $\chi_0=\chi(T_0{=}1.7\text{K}) $, $\omega_c=eB/m^*$ is the cyclotron frequency, $\hbar$ is the Planck constant, and $k_B$ the Boltzmann constant. Figure \ref{fig:QUANT}(d) shows the temperature dependence of the oscillation amplitude at $B=3.18$ T, before spin-splitting, normalized to the amplitude at $T_0=1.7$ K. By fitting the data to Eq.\ \ref{eq:effmass} we obtain an effective mass of $m^*=0.19m_e$, where $m_e$ is the free-electron mass, and a transport lifetime $\tau_t=\mu m^*/e = 1.06$ ps. The $m^*$ value is in agreement with measurements performed on natural Si \cite{fang1977effective} and corresponds to the expected value obtained from band structure calculations neglecting many-body effects \cite{ando1982electronic}.

Once the effective mass is measured, the quantum lifetime $\tau_q$ can be determined from the SdH oscillation envelope at $T_0$, using the relation \cite{bauer_1972,celik_2011}
\begin{equation}
\Delta\rho_{xx}(T_0,B)\sim\sqrt{B}\dfrac{\chi_0}{\sinh\chi_0}\exp\left(-\dfrac{\pi}{\omega_c\tau_q}\right) \, .
\end{equation}
The Dingle plot of Fig.\ \ref{fig:QUANT}(e) reports the fit from which we extract $\tau_q=0.69$ ps. The obtained $\tau_q$ implies a Landau level broadening of $\Gamma\approx\hbar/2\tau{_q}=480$ $\mu$eV, which sets an upper bound to valley splitting at the investigated density (electric field) and magnetic field. For comparison, a valley splitting energy of 275 $\mu$eV was measured in $^{28}$Si quantum dots fabricated on the same wafer in an academic environment \cite{petit2018spin}. The electron $g$-factor is evaluated by considering that the onset of spin-splitting at $B_{S}$ implies a Zeeman energy $g\mu_BB_S\simeq\Gamma$, where $\mu_B$ is the Bohr magneton. From this, a $g$-factor of $g=1.92\pm0.07$ is estimated, which is close to the expected single-particle value of $g=2$.

In conclusion, we investigated the structural and quantum transport properties of isotopically enriched $^{28}$Si/$^{28}$SiO$_2$ stacks deposited on 300 mm wafers in an industrial CMOS fab. The material characterization shows that the level of control achieved in the growth process results in a uniform deposition with high purity epilayers and a sharp semiconductor/oxide interface. Detailed quantum transport characterization of Hall-bar devices fabricated in an academic cleanroom points to a high-quality $^{28}$Si/$^{28}$SiO$_2$ interface, promising for hosting spin qubits. Mobility and critical density for these stacks are among the best reported for equivalent oxide thicknesses, with the potential to achieve large valley splitting. Disorder at the critical semiconductor/oxide interface is expected to further decrease by processing the entire gate stack in the high volume manufacturing environment, because an advanced process control is attainable and e-beam induced damage is avoided.

\section*{Acknowledgements}
M.V. acknowledges support from the Netherlands Organisation of Scientific Research (NWO) Vidi program. L.M.K.V., M.V. and G.S. acknowledge financial support by Intel Corporation.


\providecommand{\noopsort}[1]{}\providecommand{\singleletter}[1]{#1}%

\end{document}